\documentstyle[12pt]{article}
\newcommand{\rf}[1]{(\ref{#1})}
\newcommand{\beq}{\begin{equation}}
\newcommand{\eeq}{\end{equation}}

\renewcommand{\a}{\alpha}

\newcommand{\bea}{\begin{eqnarray}}
\newcommand{\eea}{\end{eqnarray}}

\newcommand{\cS}{{\cal S}}

\newcommand{\cT}{{\cal T}}

\newcommand{\cL}{{\cal L}}

\newcommand{\oh}{\frac{1}{2}}

\newcommand{\bR}{{\bf R}}

\begin{document}
\topmargin 0pt
\oddsidemargin 5mm
\headheight 0pt
\topskip 0mm

\addtolength{\baselineskip}{0.20\baselineskip}

\pagestyle{empty}
\hfill RH-13-91

\hfill October 1991

\begin{center}

\vspace{40pt}
{\Large \bf
Intrinsic and extrinsic geometry of \\ random surfaces}

\vspace{3.3 truecm}
Thordur Jonsson \\
The Science Institute \\
University of Iceland \\
Dunhaga 3 \\
107 Reykjavik \\
Iceland
\end{center}
\vfill
{\bf Abstract.} We prove that the extrinsic Hausdorff dimension is always
greater than or equal to the intrinsic Hausdorff dimension in models of
triangulated random surfaces with action which is quadratic in the
separation
of vertices.  We furthermore derive a few naive scaling relations which
relate the intrinsic Hausdorff dimension to other critical exponents.
These relations suggest that the intrinsic
 Hausdorff dimension is infinite
if the susceptibility does not diverge at the critical point.
\vspace{3 truecm}

\newpage
\pagestyle{plain}

\section{Introduction}
The geometrical properties of random surfaces can be
studied from two points of view.  One can look at the surfaces in an
ambient imbedding space as an outside observer or one can study the surfaces
as an observer living in the two-dimensional
world defined by the surface.  The first point of view, the extrinsic one,
is appropriate in string theory and in the statistical theory of
membranes while the second point of view is the only meaningful one
in quantum gravity.  There are distinct observables and critical exponents
associated with these two viewpoints.  Some of these exponents
were evaluated for
multicritical branched polymers in \cite{old}.

In this letter we consider the gaussian dynamical triangulation model
\cite{gaussian} and prove that the intrinsic Hausdorff
dimension is always smaller than or equal to the extrinsic
Hausdorff dimension.
This is a rather simple consequence of the fact that the geodesic distance
between two points on a surface is almost
always greater than or equal to the
Euclidean distance in imbedding space between the same points.  We expect
the result to be model independent and it is certainly fulfilled in all
known examples.  The proof presented below uses however the quadratic
nature of the action in an essential way.

Recently the intrinsic Hausdorff dimension of discretized
pure 2-dimensional gravity
has been studied numerically and found to be infinite \cite{ma} (see
\cite{num1,bkkm} for related older simulations) and in perturbation
theory \cite{david} using the
Liouville formulation.  Unfortunately our inequality for
the Hausdorff dimensions has no direct bearing
on this case.  The inequality is only meaningful when the
extrinsic Hausdorff dimension is defined
and this requires $d\geq 1$.   Presumably the inequality holds for
gravity coupled to matter with  central charge $c\geq 1$.  There does not
seem to be
any good
reason to expect the inequality to be valid for the analytically continued
external Hausdorff dimension to $d<1$.

In the second half of the paper we discuss alternative definitions of the
intrinsic Hausdorff dimension and derive
naive scaling relations which relate
the intrinsic Hausdorff dimension to other critical exponents in
any imbedding
dimension.  Provided our
scaling assumptions are valid, it is hard to escape the conclusion that the
Hausdorff dimension is infinite if the critical exponent of the
susceptibility is negative.
\section{Intrinsic and extrinsic Hausdorff dimension}
Let $\cT _N$ denote the
collection of all triangulations of the sphere with $N$ vertices, one of
which is singled out and called the marked vertex.  We label the marked
vertex by $0$ and label the others arbitrarily by
$i=1,2,\ldots ,N-1$.  A random surface in $\bR ^d$ based on a given
triangulation $T\in \cT _N$ is a mapping from the vertices of $T$ into $\bR
^d$, $i\mapsto x_i$.  To each such surface we assign the gaussian action
\beq
S_T=\sum_{(ij)\in \cL (T)}(x_i-x_j)^2, \label{gaussian}
\eeq where $\cL (T)$ is the collection of all nearest neighbour pairs of
vertices in in $T$.  The canonical partition function is
defined by
\beq
Z_N=\sum_{T\in\cT _N}\rho (T)\int e^{-S_T}\prod
_{i=1}^{N-1}dx_i,
\eeq
where $\rho$ is a non-negative weight factor for triangulations of the form
discussed in \cite{adfo} and we have removed the translational degree of
freedom by fixing the vertex $i=0$ at the origin in $\bR ^d$.  The Gibbs
state associated with $Z_N$ is defined by
\begin{eqnarray}
<(\cdot )>_N & = & Z^{-1}_N \sum_{T\in\cT _N}
\rho (T)\int (\cdot ) e^{-S_T}\prod _{i=1}^{N-1}dx_i \nonumber \\
  & = & Z_N^{-1}(2\pi )^{{N-1\over 2}d} \sum_{T\in\cT _N}
\rho (T) (\cdot ) \det C_T,
\end{eqnarray}
where $C_T$ is the modified adjacency matrix of $T$, see \cite{adfo}.  We
remind the reader of the graph theoretical
result that we shall use later on:
\beq
\det C_T=\#\{ B: B \;\;\mbox{\rm is a spanning tree subgraph of }
T\}.\label{trees}
\eeq
For a proof of this see e.g. \cite{graphtheory}.

The mean square extent of a surface with $N$ vertices is defined by
\beq
R^2_N=<N^{-1}\sum_{i=0}^{N-1}x_i^2>_N.
\eeq
If there is a number $\delta _{{\rm ext}}>0$, such that asymptotically
\beq
R_N^2\sim N^{2/\delta _{{\rm ext}}},
\eeq
then we call $\delta _{{\rm ext}}$ the extrinsic Hausdorff dimension of the
random surfaces in the given model.  It is clear that $\delta _{{\rm ext}}
\geq 2$ if it exists.  If $R_N$ grows more slowly than any power of $N$,
then it is customary to say that the extrinsic Hausdorff dimension is
infinite.

The mean square extent of surfaces based on regular square
triangulations of the torus can be
evaluated and the result is
\beq
R^2_N\sim \ln N ,
\eeq
in all dimensions $d$, see e.g. \cite{crystalline}.  Branched polymers
are easily
seen to have extrinsic Hausdorff dimension 4, see e.g. \cite{old}.  In
dimension d=-2 the natural analytic continuation of $\delta _{{\rm ext}}$
can be evaluated in the case when the weight factor $\rho (T)$ is given by
the order of the symmetry group of $T$ to the power -1 \cite{bkkm}.  The
result is $\delta _{{\rm ext}}=\infty$.  A very suggestive calculation in
one dimension gives the same result \cite{1dim}.
Several numerical studies have been made
of $\delta _{{\rm ext}}$ and the results are
rather inconclusive, see \cite{num1,bkkm}, but indicate that the
extrinsic Hausdorff dimension is infinite for negative imbedding
dimensions but becomes finite around $d=2$ and may decrease to 4 at large
$d$, where one expects a branched polymer phase.  It is possible that
the extrinsic Hausdorff dimension is nonuniversal, i.e. depends on $\rho$.

Next we define the intrinsic Hausdorff dimension.  Let $i$ be a vertex in
$T\in \cT _N$ and denote by $d_T(i)$ the geodesic distance from the marked
vertex $0\in T$ to $i$, i.e. $d_T(i)$ is the smallest number $n$ of links
$l_1,\ldots l_n$ in $T$ such that $0$ is an endpoint of $l_1$, $i$ is an
endpoint of $l_n$ and $l_j$ and $l_{j+1}$ share exactly one vertex for
$j=1,2,\ldots ,n-1$.  We define the intrinsic Hausdorff dimension
$\delta _{{\rm int}}$ (if it
exists) by the asymptotic formula
\beq
<N^{-1}\sum_{i=0}^{N-1}d_T(i)^2>_N\;\sim N^{2/\delta _{{\rm int}}}.
\eeq

The intrinsic Hausdorff dimension of multicritical branched polymers was
evaluated in \cite{old}, see also \cite{david}.  In the case
of regular triangulations $\delta _{{\rm
int}}$ is trivially 2.
We shall now prove that for any choice of weight
factor $\rho$  for which the intrinsic and extrinsic Hausdorff dimensions
exist they satisfy the inequality
\beq
\delta _{{\rm int}} \leq \delta _{{\rm ext}}.\label{spiral}
\eeq

Let $i\in T$ and $n=d_T(i)$.  Let $j(\a )$, $\a =0,1,\ldots ,n$, $j(0)=0$,
$j(n)=i$,  be the
vertices in a geodesic path in $T$ from $0$ to $i$.  Then
\beq
x_i=\sum_{\a =1}^n(x_{j(\a )}-x_{j(\a -1)}),
\eeq
so
\beq
x_i^2\leq n\sum_{\a =1}^n(x_{j(\a )}-x_{j(\a -1)})^2.
\eeq
Hence,
\beq
<\sum_{i=0}^{N-1}x_i^2>_N\;\leq Z_N^{-1}\sum_{T\in\cT _N}\rho (T)\sum
_{i=0}^{N-1}d_T(i)\sum_{\a =1}^{d_T(i)}\int (x_{j(\a )}-x_{j(\a -1)})^2
 e^{-S_T}
\prod_{i=1}^{N-1}dx_i.
\eeq
Now we use that
\beq
(x_{j(\a )}-x_{j(\a -1)})^2 e^{-S_T}\leq K e^{-S_T^{\prime}},
\eeq
where $K$ is a constant and the modified action $S_T^{\prime}$
is defined by leaving out the term $(x_{j(\a )}-x_{j(\a -1)})^2$
in the action \rf{gaussian}, i.e.
\beq
S_T^{\prime}=\sum_{(k,l)\in \cL (T^{\prime})}(x_k-x_l)^2,
\eeq
and here $T^{\prime}$ is the graph obtained by removing the link
$(j(\a ),j(\a -1))$ from $T$.
 Note
that $T^{\prime}$ is not a triangulation, but the formula \rf{trees}
is still
valid for $\det C_{T^{\prime}}$.  Any tree that spans $T^{\prime}$ also
spans $T$, so
\beq
\det C_{T^{\prime}}\leq  \det C_T.
\eeq
We furthermore claim that
\beq
\det C_{T}\leq  3\det C_{T^{\prime}}.
\eeq
Hence,
\beq
R^2_N\leq K^{\prime} N^{-1}<\sum_{i=0}^{N-1}d_T(i)^2>_N,
\eeq
where $K^{\prime}$ is another constant
and the inequality \rf{spiral}
follows for any imbedding dimension $d$.

In order to prove the claim we note that the trees that span $T$ fall into
two disjoint classes: those which contain the link $\lambda =(j(\a -1),j(\a
))$ and the ones that do not contain $\lambda$.  The trees in the second
class are precisely those that span $T^{\prime}$.  Consider a tree $B$ in
the first class.  Let $k$ be the third vertex in one of the two triangles
in $T$ that share the link $\lambda$.  If we remove $\lambda$ from $B$, we
obtain two disjoint trees $B_1$ and $B_2$ with $j(\a -1)\in B_1$ and $j(\a
)\in B_2$.  Let us assume that $k\in B_1$.  Define a tree $B^{\prime}$ by
removing the link $\lambda$ from $B$ and replacing it with the link
$(k,j(\a ))$.  If $k\in B_2$, we define $B^{\prime}$ by replacing $\lambda$
with $(k,j(\a -1))$.  In both cases $B^{\prime}$ is a spanning tree of
$T^{\prime}$ and the desired result follows, since the mapping $B\mapsto
B^{\prime}$ is at most two to one.

\section{Scaling relations for intrinsic exponents}
The above definition of the Hausdorff dimension is not the only possible
one.  It is not unreasonable to define a Hausdorff dimension in the grand
canonical ensemble relating the average area of surfaces to a diverging
correlation length as the critical point is approached.  One can also, in
the case of the intrinsic Hausdorff dimension, consider the volume $V$ of a
ball of radius $n$ and use the relation between the average value of $V$
and $n$ to define a Hausdorff dimension.  It is by no
means clear that the two intrinsic
Hausdorff dimensions defined in this way in the grand canonical
ensemble are the same
or identical to the dimension defined in the canonical ensemble in the
previous section.
Here we shall study the two definitions of the intrinsic Hausdorff dimension
in the grand canonical ensemble and give a naive derivation
of their relation to other intrinsic critical exponents.

Let $\cS _{2,n}$ be the collection of all triangulations with two marked
vertices a distance $n$ apart.  The intrinsic two point function
$G_{\mu}(n)$ is defined by
\beq
G_{\mu}(n)=\sum _{T\in\cS _{2,n}}e^{-\mu |T|}W(T),
\eeq
where $W(T)$ is a weight factor coming from $\rho (T)$ and the "matter
fields":
\beq
W(T)=\rho (T)\int e^{-S_T}\prod _{i=1}^{|T|-1}dx_i.
\eeq
Here $|T|$ denotes the number of vertices in $T$.  With this definition of
$W$ there is a $\mu _c>0$, such that the two point function is finite for
$\mu >\mu _c$ and the sum defining $G_{\mu}(n)$ diverges for $\mu <\mu _c$
\cite{adfo}.  The arguments presented below do not make any use of the
detailed form of $W$ and are valid for the analytic continuation of $W$
to $d<1$.

The susceptibility is defined by
\beq
\chi (\mu )=\sum _{n=1}^{\infty}G_{\mu }(n)
\eeq
and the partition function by
\beq
Z(\mu )=\sum _{T\in \cS _1}e^{-\mu |T|},
\eeq where $\cS _1$ is the collection of all triangulations with one marked
vertex.  If $\chi$ diverges at $\mu _c$, we assume that there is $\gamma
>0$ such that $\chi (\mu )\sim (\mu -\mu _c)^{-\gamma}$, but if $\chi (\mu
_c)<\infty $, then we assume that
$\chi (\mu _c)- \chi (\mu )\sim (\mu -\mu _c)^{-\gamma}$
for some $\gamma <0$.

By subadditivity arguments, see \cite{dfj}, one can show that
\beq
G_{\mu }(n)\sim e^{-m(\mu )n},
\eeq
where $m(\mu )\geq 0$.  We assume that
\beq
m(\mu )\sim (\mu -\mu _c)^{\nu}
\eeq
for some $\nu >0$ as $\mu\to\mu _c$,
and define the correlation length $\xi$
by
\beq
\xi (\mu)=m(\mu )^{-1}.
\eeq
Assuming that triangulations with linear size $\xi$ give a dominating
contribution to the two point function and assuming that there is $\delta
_1 > 0$ such that
\beq
	G_{\mu}(\xi)^{-1}\sum _{T\in \cT _{2,\xi}}|T|e^{-\mu |T|}W(T)\sim \xi
^{\delta _1},
\eeq
it follows that
\beq
\delta _1\nu=1.\label{s1}
\eeq

If $T\in\cS _1$, let $D_T(n)$ be the number of points in $T$ at a distance
$n$ from the marked point.  Assume that there is a $\delta _2\geq 0$ such
that
\beq
Z^{-1}\sum _{T\in \cT _1}D_T(\xi )e^{-\mu |T|}W(T)\sim \xi ^{\delta _2-1}.
\label{1}
\eeq
The numbers $\delta _1$ annd $\delta _2$ are the two candidates for an
intrinsic Hausdorff dimension.

Let us next assume that
\beq
G_{\mu}(\xi )\sim \xi ^{-\eta }\label{2}
\eeq
and the scaling limit of $G_{\mu}$ exists.  Then we obtain a Fisher scaling
relation, which in the present context takes the form
\beq
\nu (1-\eta )=\max \{ 0,\gamma \}\label{fisher}
\eeq
by the same proof as in \cite{dfj2}.

Finally note that if $Z$ is finite at the critical point, then
\beq
\eta = 1-\delta _2 \label{s2}
\eeq
by Eqs. \rf{1} and \rf{2}, since
\beq
\sum _{T\in \cT _1}D_T(\xi )e^{-\mu |T|}W(T)=G_{\mu}(\xi ).
\eeq
Suppose now $\gamma <0$.  Then
\beq
\nu (1-\eta )=0,
\eeq
so $\eta =1$ or $\nu =0$.  If $\eta =1$, then $\delta _2=0$ which
presumably does not happen in any sensible model.  We conclude that $\nu
=0$ which implies that the Hausdorff dimension $\delta _1$ is infinite.
If we in addition assume that $\delta _1=\delta _2$, then $\eta =-\infty$.

If $\gamma >0$ and we assume that $\delta _1=\delta _2$, then we can solve
equations \rf{s1}, \rf{fisher} and \rf{s2} for $\gamma$ and the result is
\beq
\gamma =1.
\eeq
This, however, is impossible since the inequality $\gamma \leq \oh $ holds
in all models of the type we are considering \cite{adfo,dfj}.  We conclude
that the Hausdorff dimensions $\delta _1$ and $\delta _2$
must be different or at least one of the scaling assumptions is not valid.

For ordinary branched polymers one finds \cite{old} $\gamma =\oh $,
$\nu =\oh $ and $\eta =0$ by a direct calculation and hence $\delta _1=2$
and $\delta _2=1$ by the scaling relations.  In this case
$\delta _{{\rm int}}=2$ as remarked in
\cite{david}, which throws some doubt on the validity of the scaling
relation
\rf{s2}.

\section{Discussion}
The intrinsic
Hausdorff dimension calculated in \cite{ma} is defined in almost
the same
way as $\delta _2$ but in a different ensemble.  It is therefore possible
that the scaling relations in the last section and the fact
that $\gamma <0$
for pure 2d-gravity explain the numerical
results of \cite{ma}.  In fact, all
numerical evidence is consistent with $\gamma <0$ and $\delta _{{\rm int}}
=\infty$ for $d<1$, since $\delta _{{\rm ext}}$ seems to be infinite for
$d\leq 1$.

One must bear in mind that the exponents that we have considered here
give a very incomplete picture of what a typical surface looks like.
This is best illustrated by the fact that $\delta _{{\rm int}}=2$ both for
ordinary branched polymers and for surfaces with a regular triangulation.

An observable of great interest is the number of connected components of
the boundary of a ball of radius $n$ as $n$ gets large.  This quantity was
studied numerically in \cite{ma} and found to increase rapidly with $n$.
This means that in some sense the surfaces of 2d-gravity are similar to
branched polymers.  However, these surfaces are not ordinary branched
polymers, since they have susceptibility exponent $\gamma =-\oh$, whereas
$\gamma =\oh$ for branched polymers with positive weight.
A clarification of this issue would be of utmost importance.

\medskip

\medskip
\noindent
{\bf Acknowledgement.} I would like to thank
Jan Ambj\o rn, Bergfinnur Durhuus, Fran\c{c}ois David and
John Wheater for helpful discussions and Krszystof Gawe\c{d}zki
for hospitality
at IHES.  I am furthermore indebted to the
theoretical physics group at
Imperial College for hospitality.

\end{document}